# The detection of *Fermi* AGN above 100 GeV using clustering analysis

Thomas Armstrong,★ Anthony M. Brown, Paula M. Chadwick and S. J. Nolan
*Department of Physics and Centre for Advanced Instrumentation, Durham University, Durham, DH1 3LE, UK*

**ABSTRACT**

The density-based clustering algorithm DBSCAN has been applied to the *Fermi* Large Area Telescope (LAT) data set of $E_\gamma \geq 100$ GeV events with $|b| > 10°$, in order to search for new very high energy (VHE) $\gamma$-ray sources. The clustering analysis returned 49 clusters, of which 21 correspond to already known VHE-emitting active galactic nuclei (AGN) within the TeVCat catalogue and a further 11 were found to be significant in a full *Fermi* analysis. Of these, two are previously detected *Fermi* VHE AGN, and nine represent new VHE sources consisting of six BL Lac objects, one blazar of unknown type and two unassociated sources. Comparing these, along with the VHE AGN RBS 0679 and RBS 0970 previously detected with *Fermi*-LAT, to the current populations of AGN detected with ground-based instruments and *Fermi* suggests that the VHE-emitting AGN discovered in this study are very similar to the TeVCat AGN and therefore further observations with ground-based imaging atmospheric Cherenkov telescopes are recommended.

## 1 INTRODUCTION

Since its launch in 2008, the *Fermi* space-based $\gamma$-ray telescope has spent ∼95 per cent of its time in *all-sky-survey* mode, in which the Large Area Telescope (LAT) scans the entire sky every two orbits, or approximately every 3 h (Atwood et al. 2009). Information on location, time and energy is recorded for each event detected, resulting in a large, multidimensional data base which provides us with a wealth of information about the $\gamma$-ray sky.

The *Fermi* 2 year point source catalogue (2FGL), which was released in 2012 (Ackermann et al. 2012a), was the main source of information for $\gamma$-ray sources until the recent release of the 3rd source catalogue (3FGL; Acero et al. 2015). The method used to create this data set relied on wavelet-based algorithms, such as *mr_filter* (Starck & Pierre 1998) and *PGWave* (Damiani et al. 1997; Ciprini et al. 2007), minimum spanning trees (MST; Campana et al. 2008) and the addition of *pointlike* (Kerr 2010) in the 3FGL to find 'seeds. These were all then followed up with a full likelihood analysis. However, in the last two decades and over many disciplines, there has been a substantial amount of work on clustering analysis as a major statistical technique for classifying large data sets into meaningful subsets. It is therefore likely that these methods are worthy of investigation as potential source-finding algorithms for the LAT data set.

Alongside the MST clustering performed in the source detection for the 2FGL, investigation into clustering performance for *Fermi* was carried out in Tramacere & Vecchio (2013) using the density-based clustering algorithm DBSCAN (Density-Based Spatial Clustering of Applications with Noise; Ester et al. 1996). By applying the cluster analysis to simulated *Fermi*-LAT data, Tramacere & Vecchio were able to show the statistical robustness of the code's ability to identify potential sources in noisy regions.

In this paper, we have chosen to apply a cluster analysis to all $E_\gamma \geq 100$ GeV photons with $|b| > 10°$. First, since the extragalactic diffuse background has a spectral index of 2.41, we reduce complications due to background noise which mainly affect lower energies (Abdo et al. 2010). Secondly, as the computational complexity of DBSCAN runs as $O(n^2)$,[1] by using only the high-energy events we are able to run a full, unbiased and model-independent clustering analysis of the whole sky without using a large amount of computing time. Finally, the possibility of increasing the known very high energy (VHE) $\gamma$-ray population of TeVCat[2] active galactic nuclei (AGN) from its current number of 61 is attractive, particularly in the light of framing the scientific priorities for the forthcoming Cherenkov Telescope Array (CTA; Actis et al. 2011; Sol el al. 2013).

This paper is organized as follows. In Section 2, we discuss the use of clustering, the chosen method and its application to *Fermi*-LAT VHE events. The clusters are verified using *Fermi* tools in Section 3 and the results, along with a preliminary analysis of the global properties of the detected sources, are discussed in Section 4.

---

★ E-mail: thomas.armstrong@durham.ac.uk

[1] It is possible to improve the speed up to $O(n \log n)$ by pre-computing the EPS-neighbourhoods (see Section 2.1). However, the computational demand of this work did not require this step.
[2] Online catalogue of VHE ground-based detections http://tevcat.uchicago.edu/ (Horan & Wakely 2008).

## 2 CLUSTERING ALGORITHMS

There are many methods that fall under the classification of clustering algorithms. However, they can usually be divided into three main types: partitioning, which is based on iterative relocation of data points between clusters and generally requires advance knowledge of the number of clusters (e.g. K-Means; MacQueen 1967); hierarchical, which groups data with a sequence of nested partitions by iteratively splitting the data base into smaller subsets, using either a top-down or bottom-up approach, and density-based which produces clusters if the number of events within a certain area are greater than the number in its surroundings, either through density-connected points or based on an explicitly defined density function. One of the simplest and most widely-used examples of this is DBSCAN (Ester et al. 1996). Its ability to pick out clusters of arbitrary shape from noisy data and its use in previous preliminary studies (Carlson et al. 2013; Tramacere & Vecchio 2013) make it the logical choice for examining clustering in the *Fermi*-LAT data.

### 2.1 DBSCAN

The base version of DBSCAN requires two input parameters, MinPts, the smallest number of events we would consider to constitute a cluster within a circle of radius EPS, which is the second parameter. The main concept of DBSCAN centres around the idea of *core samples* in areas of higher density. A core sample is defined as a point, $p$, which satisfies the condition $N_{\rm EPS}(p) \geq {\rm MinPts}$. That is, $p$ is a core point if the number of events within its EPS-neighbourhood is equal to or greater than that given by the MinPts parameter.

The code we used is built on the Scikit–Learn PYTHON library (Pedregosa et al. 2011) in which the clusters are computed as follows:

(i) for each point $p$ in a set of objects $D$, the number of points within the EPS-neighbourhood ($N_{\rm EPS}(p)$) is found;
(ii) if the core sample condition $N_{\rm EPS}(p) \geq {\rm MinPts}$ is satisfied, then $p$ is a core point and is added to the cluster $C$;
(iii) if a point $q$ within the EPS-neighbourhood of the core point $p$ also satisfies the core sample condition then $p$ and $q$ are *density-connected* and $q$ is added to $C$. If not, it is classified as a *border point* or *density-reachable*;
(iv) step (iii) is repeated for every candidate core point for $C$;
(v) the algorithm moves to a new, unprocessed, core point and returns to step (ii).

In brief, the DBSCAN algorithm takes every point, considers whether it is in a dense region and then builds up a cluster by adding all nearby points, with respect to EPS and MinPts, that exist above a certain density. All objects that have been processed but are not considered density-connected to a cluster are defined as noise. This allows us to build clusters of arbitrary shape and to reject background efficiently (see Fig. 1).

One of the difficulties of using DBSCAN is that the initial choice of EPS and MinPts strongly affect the outcome of the clustering algorithm. However, Tramacere & Vecchio (2013) performed a statistical analysis using DBSCAN on simulated data to determine optimum choices for EPS and MinPts and to test the robustness of the algorithm. One conclusion that was drawn is that EPS can be related to the point spread function (PSF) of the LAT detector. In the case of our application to clustering above 100 GeV, as the *Fermi*-LAT response functions give a PSF of $0°.12$ at 100 GeV for a 68 per cent containment radius and $0°.5$ for 95 per cent (Ackermann et al. 2012b), we investigated a range between these values.

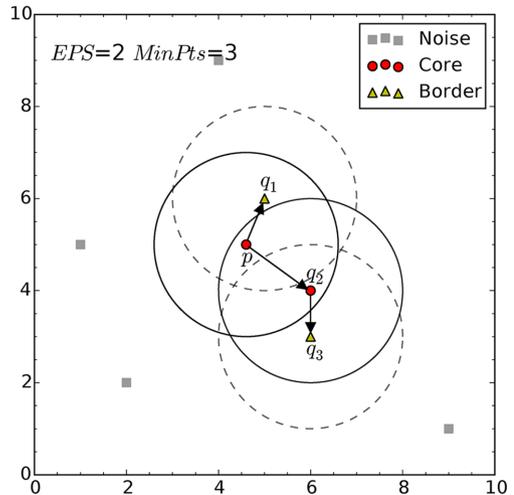

**Figure 1.** Description of DBSCAN with EPS=2 MinPts=3. Starting with point $p$, which is classified as a core point as there is a total of three points within the *EPS*-neighbourhood, $q_1$, $q_2$ and $p$ itself. $q_1$ is directly density-reachable from $p$ but does not itself satisfy the core condition and is therefore defined as a border point. $q_2$ is also directly density-reachable from $p$ but is a core point. the classification of $q_3$ follows that of $q_1$. All the rest of the points are considered as noise. The resulting cluster has two core points and a total of four if including border points.

A second limitation of DBSCAN is its inability to deal with a spatially non-uniform background. In these cases, the intrinsic cluster structure may be masked by a non-ideal global set of parameters. For example, we may fail to reveal substructure in areas of general over density in favour of finding clusters in fainter regions. Conversely, we may sacrifice these more diffuse clusters in order to obtain a characterization of the bright area. There are clustering algorithms available (e.g. OPTICS and ENDBSCAN; Ankurst et al. 1999; Roy & Bhattacharyya 2005) that modify DBSCAN to allow for its application to data with variable noise. Alternatively one could run DBSCAN in a scanning mode, adjusting the input parameters for each scan region, which was the approach taken in both Tramacere & Vecchio (2013) and Carlson et al. (2013). For our application of clustering off-plane at energies greater than 100 GeV, the variation in the diffuse background is greatly reduced to the point where it can be considered negligible.

### 2.2 Clustering of VHE γ-ray events

The VHE domain provides a good test bed for the validation of DBSCAN. By restricting ourselves to energies greater than 100 GeV, we not only reduce the problem of varying background, but also the computational power needed to perform an unbiased clustering search of the whole sky. With its long exposure time and full sky coverage, *Fermi* gives us access to the deepest extragalactic scan presently available at these energies. Indeed, recent work took advantage of *Fermi*-LAT's deep exposure to discover two new VHE-bright AGN (Brown 2014; Brown, Chadwick & Landt 2014). It is important to note, however, that these studies only searched for VHE emission around bright, spectrally hard, *Fermi*-LAT detected BL Lac objects. Given the relatively small number known of VHE γ-ray objects, it is important that we investigate statistical methods in the context of a model-independent search, which could lead to greater understanding of VHE populations.

For our data set, we took all *Fermi*-LAT events for the first 6.25 years of operation from 2008 August 04 to 2014 November

**Table 1.** Selection criteria for the data to which the clustering analysis was applied.

| Cut name | Value |
| --- | --- |
| Tools version | v9r31p1 |
| Response function | pass 7 repo |
| Emin | 100 GeV |
| Emax | 300 GeV |
| $T_{\text{start}}$ (MET) | 239 557 417 |
| $T_{\text{end}}$ (MET) | 438 847 466 |
| Zenith | 100° |
| Evclass | Source (2) |
| Conversion type | Front & back |
| DATA_QUAL | 1 |
| LAT_CONFIG | 1 |
| ABS(ROCK_ANGLE) | <52° |

28 (Mission Elapsed Time: 239 557 417 to 438 847 466) and selected events with energies of greater than 100 GeV for both front- and back-converting SOURCE class events. We also excluded the Galactic plane ($|b| < 10°$) from our scan as the source confusion resulting from the poor angular resolution prevents us from reliably picking out individual clusters in this dense region.

In accordance with the pass 7 rep criteria, a zenith cut of 100° was applied to the data to remove any $\gamma$-rays induced by cosmic ray interactions in the Earth's atmosphere. The good time intervals were generated by applying a filter expression of '(data qual == 1) && (lat config == 1) && abs(rock angle) <52°' to the data, where the (data qual) and (lat config) flags remove suboptimal data affected by spacecraft events and the (abs(rock angle)) flag removes data periods where the LAT detector rocking is greater than 52°. These criteria are summarized in Table 1.

For our clustering parameters, based on Tramacere & Vecchio (2013) we chose to investigate a range of EPS values between the 68 per cent and 95 per cent containment radii. Using the pass 7 response files for the PSF at 100 GeV for both front- and back-converting events, we found this to equate to a range of ∼0°.12 to 0°.5. As we are considering relatively low statistics, we chose MinPts to be the minimum number of events that could constitute a cluster statistically, namely three events.

For each cluster, the effective radius from the cluster centroid was calculated as $r_{\text{eff}} = \sqrt{\sigma_x^2 + \sigma_y^2}$, where $\sigma_x$ and $\sigma_y$ are the uncertainties expressed as the standard deviations in the event position. To determine the significance of the cluster we applied the Likelihood Ratio Test (LRT) as described in Li & Ma (1983) and applied in both Tramacere & Vecchio (2013) and Carlson et al. (2013),

$$s = \sqrt{2\left(N_{\text{s}} \ln\left[\frac{2N_{\text{s}}}{N_{\text{s}} + N_{\text{b}}}\right] + N_{\text{b}} \ln\left[\frac{2N_{\text{b}}}{N_{\text{s}} + N_{\text{b}}}\right]\right)}, \quad (1)$$

where $N_{\text{s}}$ is the number of events taken from the DBSCAN and includes core and border events. The background $N_{\text{b}}$ was estimated from the number of events between $2r_{\text{eff}}$ and $3r_{\text{eff}}$. We set a cluster significance of $s = 2$ as our minimum significance for a cluster. When $N_{\text{s}}$ and $N_{\text{b}}$ are large, which is not the case here, this represents a fluctuation of $2\sigma$ above the background. Therefore we use the LRT only as an indicator and in Section 4.1 we discuss the validity of this assumption.

A study of the effects of changing EPS, described in Section 4.1, has shown the optimal value to be 0°.4. The results from the cluster analysis for sources with an LRT significance of s > 2 using an EPS of 0°.4 can be found in Tables 2 and 3, where Table 2 lists the

28 sources that are not currently part of the TeVCat VHE catalogue and Table 3 lists the further 21 that are and have been included for reference.

## 3 VERIFICATION OF VHE CLUSTERS USING *Fermi* ANALYSIS

For each significant cluster found using the DBSCAN algorithm, we used the full 6.25 years' worth of the *Fermi*-LAT data within an ROI of radius 5° surrounding the cluster position for further analysis. As before, the data were reduced with the *Fermi* tools gtselect and gtmktime in order to apply a zenith cut and to keep only the 'good time intervals' according the same pass 7 criteria for SOURCE class events between 100 and 300 GeV (see Table 1).

We ran an unbinned likelihood analysis on each source, modelling each cluster with a power-law spectral shape of the form,

$$\frac{dN}{dE} = A \times \left(\frac{E}{E_o}\right)^{-\Gamma}, \quad (2)$$

where $A$ is the normalization, $\Gamma$ the spectral index and $E_o$ the scaling factor. In addition to modelling the cluster, each analysis used a model file consisting of all point sources within 15° of the cluster position, as well as the most recent Galactic and extragalactic diffuse models (gll_iem_v05_rev1.fit and iso_source_v05.txt, respectively). The position and the spectral shape of these point sources were taken from the third *Fermi* catalogue (3FGL; Acero et al. 2015). Furthermore, several clustering events were found to be located in close proximity to known extended $\gamma$-ray sources, namely W28, W30, W44, the Cen A lobes and HESS J1841-055. These extended sources were accounted for with their respective spatial distribution models from the 3FGL. During the analysis, the normalization and the spectral index of the cluster source and the point sources within the ROI where left free to vary. Modelled sources outside the ROI but within 15° had their parameters frozen to those published in the 3FGL.[3] Likewise, the normalization factor of the extragalactic diffuse emission was left free to vary, and the Galactic diffuse template was multiplied by a power law in energy, the normalization of which was left free to vary (Ackermann et al. 2012c).

From the unbinned analysis with the above model, we arrived at a best-fitting power-law model and integrated flux for each cluster along with resulting likelihood Test Statistic (TS).[4] If the analysis returned an insignificant result (TS <25) for the $E_\gamma \geq 100$ GeV flux, upper limits to the flux were calculated.

To ensure that there were no $\gamma$-ray sources in the ROI that were not taken into account in the model (such as other sources not included in the 3FGL), we used the *Fermi* tool gttsmap and the best-fitting model to create a TS significance map. The figures available in the supplementary online material show the TS value distribution within the ROI of all sources found with TS> 25. Apart from the central source there is no other significant source within the field of view, with two exceptions: 1ES 1011+496, which is located 2°.89 away from 3FGL J1031.2+5053, and Markarian 421

---

[3] In some cases extra parameters were frozen in order to improve the global fit. Sources with a significance less than 5 had their parameters frozen, sources with a TS <1 were removed altogether.
[4] The Test Statistic is defined as $TS = -2\ln(L_{\text{max}, 0}/L_{\text{max}, 1})$, where $L_{\text{max}, 0}$ is the maximum likelihood value for a model without an additional source (the 'null hypothesis') and $L_{\text{max}, 1}$ is the maximum likelihood value for a model with the additional source at a specified location.

**Table 2.** Results for sources detected at $E \geq 100$ GeV with DBSCAN. 'Unkn.' refers to sources that are not present in the 3FGL. The TS, flux and $\Gamma$ were found with follow-up analysis using the published *Fermi* tools. The first 11 sources are those that were found to be significant (TS >25) with the follow-up analysis. For sources with TS <25, upper limits were calculated for the flux. A binned likelihood analysis has also been applied to the energy range $0.1 > E > 100$ GeV in order to obtain a power-law spectral index, see Section 4.2.

| | *Fermi* ID | Counterpart ID | RA (°) | Dec. (°) | $z$ | $s_{0.4}$ | TS 100–300 GeV | Flux 100–300 GeV $\times 10^{-11}$ ph cm$^{-2}$ s$^{-1}$ | TS 0.1–100 GeV | Flux 0.1–100 GeV $\times 10^{-9}$ ph cm$^{-2}$ s$^{-1}$ | $\Gamma$ 0.1–100 GeV |
|---|---|---|---|---|---|---|---|---|---|---|---|
| 1 | 3FGL J0209.4-5229 | RBS 285 | 32.45 | −52.48 | – | 2.04 | 37.08 | 1.56 ± 0.74 | 690.92 | 7.56 ± 1.00 | 1.74 ± 0.053 |
| 2 | 3FGL J0543.9-5531 | RBS 0679[a] | 85.99 | −55.55 | 0.273 | 2.35 | 51.12 | 2.07 ± 0.96 | 722.77 | 8.16 ± 1.06 | 1.72 ± 0.051 |
| 3 | 3FGL J0912.9-2104 | MRC 0910-208 | 138.31 | −21.09 | 0.198 | 2.35 | 36.09 | 2.34 ± 1.09 | 278.69 | 6.25 ± 1.47 | 1.83 ± 0.085 |
| 4 | 3FGL J1031.2+5053 | RBS 877 | 157.74 | 50.88 | 0.360 | 2.04 | 27.97 | 1.59 ± 0.89 | 465.99 | 5.39 ± 0.030 | 1.77 ± 0.0024 |
| 5 | 3FGL J1117.0+2014 | RBS 958 | 169.24 | 20.25 | 0.138 | 2.04 | 36.21 | 1.94 ± 1.11 | 802.22 | 14.39 ± 0.38 | 1.95 ± 0.010 |
| 6 | 3FGL J1120.8+4212 | RBS 0970[a] | 170.16 | 42.26 | 0.390 | 2.35 | 34.34 | 2.18 ± 1.13 | 730.57 | 4.31 ± 0.53 | 1.55 ± 0.050 |
| 7 | 3FGL J2322.5+3436 | TXS 2320+343 | 350.63 | 34.60 | 0.098 | 2.35 | 41.82 | 2.13 ± 1.08 | 76.50 | 2.12 ± 0.17 | 1.77 ± 0.025 |
| 8 | 3FGL J2356.0+4037 | GB6 B2353+4020 | 359.17 | 40.66 | 0.331 | 2.04 | 27.69 | 1.55 ± 0.91 | 91.68 | 2.04 ± 0.23 | 1.72 ± 0.040 |
| 9 | 3FGL J1714.1-2029 | 1RXS J171405.2-202747 | 258.48 | −20.41 | – | 2.35 | 27.34 | 2.01 ± 1.11 | 43.06 | 1.13 ± 0.88 | 1.59 ± 0.23 |
| 10 | Unkn. J2132.43-3416 | – | 323.21 | −34.24 | – | 2.04 | 25.63 | 2.45 ± 1.84 | 3.83 | <0.42 | – |
| 11 | 3FGL J2209.0-0450 | – | 332.44 | −4.86 | – | 2.04 | 25.59 | 2.60 ± 1.40 | 27.39 | 1.37 ± 0.032 | 1.80 ± 0.0078 |
| 1 | 3FGL J0730.5-6606 | PMN J0730-6602 | 112.80 | −66.00 | 0.106 | 2.04 | 19.17 | <2.10 | 102.93 | 2.59 ± 0.96 | 1.71 ± 0.13 |
| 2 | 3FGL J1309.3+4304 | B3 1307+433 | 197.21 | 42.83 | 0.690 | 2.04 | 20.10 | <1.14 | 1123.02 | 15.25 ± 0.23 | 1.92 ± 0.0067 |
| 3 | 3FGL J1659.0-0142 | – | 255.23 | −1.44 | – | 2.04 | 15.39 | <0.86 | 86.66 | 10.44 ± 3.79 | 2.16 ± 0.13 |
| 4 | 2FGL J1721.5-0718c | – | 260.18 | −7.20 | – | 2.04 | 12.95 | <4.81 | 5.17 | <13.2 | – |
| 5 | 3FGL J1838.8+4802 | GB6 J1838+4802 | 279.68 | 48.01 | 0.300 | 2.04 | 13.57 | <0.91 | 828.92 | 10.23 ± 1.06 | 1.79 ± 0.041 |
| 6 | Unkn. J0255.43+3334 | – | 43.90 | 33.57 | – | 2.04 | 16.53 | <1.20 | ~0 | <0.068 | – |
| 7 | Unkn. J0808.43+1645 | – | 122.19 | 16.75 | – | 2.04 | 18.34 | <5.08 | 0.03 | <2.16 | – |
| 8 | Unkn. J1359.3-4019 | – | 209.86 | −40.32 | – | 2.04 | 22.44 | <1.56 | 0.68 | <21.3 | – |
| 8 | Unkn. J1526.16-0515 | – | 231.57 | −5.26 | – | 2.04 | 12.36 | <1.01 | ~0 | <0.070 | – |
| 10 | Unkn. J1626.7-0617 | – | 246.73 | −6.29 | – | 2.04 | 23.45 | <1.69 | ~0 | <0.011 | – |
| 11 | Unkn. J1655.52+0052 | – | 253.99 | −0.88 | – | 2.04 | 14.19 | <1.22 | 35.37 | 18.90 ± 0.042 | 2.76 ± 0.00039 |
| 12 | Unkn. J1902.14+4557 | – | 285.41 | 46.06 | – | 2.04 | 15.53 | <2.24 | 0.67 | <0.47 | – |
| 13 | Unkn. J1903.33+3649 | – | 285.90 | 36.82 | – | 2.04 | 10.47 | <1.26 | 7.13 | <63.0 | – |
| 14 | Unkn. J1907.07-2930 | – | 286.69 | −29.36 | – | 2.04 | 10.34 | <0.49 | 1.82 | <10.9 | – |
| 15 | Unkn. J1938.09-0350 | – | 294.55 | −3.84 | – | 2.04 | 12.35 | <1.51 | 1.20 | <25.9 | – |
| 16 | Unkn. J2001.5+0330 | – | 300.47 | 3.68 | – | 2.04 | 16.44 | <1.04 | ~0 | <0.065 | – |
| 17 | Unkn. J2212.19+8221 | – | 333.08 | 82.36 | – | 2.04 | 20.43 | <1.12 | ~0 | <0.056 | – |

*Notes.* [a]The two sources RBS 0679 and RBS 0970 were discovered as VHE sources in Brown et al. (2014) and Brown (2014), respectively, but are not in the TeVCat catalogue.

**Table 3.** 21 Sources found at $E \geq 100$ GeV with DBSCAN which are also in the TeVCat and 3FGL catalogues. Here we show the 3FGL and TeVCat identifiers, the number of events found with DBSCAN, the LRT significance returned and the TS and flux from the likelihood fit.

| | *Fermi* ID | Counterpart ID | $n_{0.4}$ | $s_{0.4}$ | TS 100–300 GeV | Flux 100–300 GeV $\times 10^{-11}$ ph cm$^{-2}$ s$^{-1}$ |
|---|---|---|---|---|---|---|
| 1 | 3FGL J0222.6+4301 | MAGIC J0223+403 | 11 | 3.91 | 133.68 | 6.11 ± 1.77 |
| 2 | 3FGL J0303.4-2407 | PKS 0301-243 | 8 | 3.33 | 76.80 | 4.20 ± 1.59 |
| 3 | 3FGL J0319.8+1847 | RBS 413 | 3 | 2.04 | 24.37 | <2.13 |
| 4 | 3FGL J0319.8+4130 | NGC 1275 | 3 | 2.04 | 24.75 | <2.06 |
| 5 | 3FGL J0449.4-4350 | PKS 0447-439 | 7 | 3.12 | 67.65 | 3.47 ± 1.34 |
| 6 | 3FGL J0508.0+6736 | 1ES 0502+675 | 13 | 4.12 | 161.72 | 5.75 ± 1.50 |
| 7 | 3FGL J0650.7+2503 | 1ES 0647+250 | 8 | 3.33 | 64.54 | 4.01 ± 1.54 |
| 8 | 3FGL J0721.9+7120 | S5 0716+714 | 5 | 2.63 | 39.15 | 2.03 ± 0.91 |
| 9 | 3FGL J0809.8+5218 | 1ES 0806+524 | 3 | 2.04 | 31.85 | 1.32 ± 0.77 |
| 10 | 3FGL J1015.0+4925 | 1ES 1011+496 | 13 | 4.02 | 153.57 | 6.86 ± 1.82 |
| 11 | 3FGL J1104.4+3812 | Markarian 421 | 95 | 11.53 | 1259.24 | 50.0 ± 4.97 |
| 12 | 3FGL J1136.6+7009 | Markarian 180 | 5 | 2.63 | 31.05 | 1.14 ± 0.67 |
| 13 | 3FGL J1217.8+3007 | 1ES 1215+303 | 5 | 2.63 | 28.96 | 2.15 ± 1.11 |
| 14 | 3FGL J1221.3+3010 | 1ES 1218+304 | 9 | 3.53 | 83.24 | 4.79 ± 2.57 |
| 15 | 3FGL J1224.9+2122 | 4C 21.35 | 3 | 2.04 | 30.09 | 1.66 ± 0.98 |
| 16 | 3FGL J1427.0+2347 | PKS 1424+240 | 9 | 3.53 | 81.44 | 4.64 ± 1.64 |
| 17 | 3FGL J1555.7+1111 | PG 1553+113 | 27 | 6.08 | 287.47 | 14.3 ± 2.83 |
| 18 | 3FGL J1653.9+3945 | Markarian 501 | 47 | 8.11 | 502.40 | 22.4 ± 3.34 |
| 19 | 3FGL J2000.0+6509 | 1ES 1959+650 | 9 | 3.53 | 52.76 | 3.85 ± 1.30 |
| 20 | 3FGL J2009.3-4849 | PKS 2005-489 | 9 | 3.53 | 74.93 | 4.14 ± 1.48 |
| 21 | 3FGL J2158.8-3013 | PKS 2155-304 | 21 | 5.31 | 218.82 | 12.4 ± 2.75 |

which is 5°.08 away from 3FGL J1120.8+4212. However these have been accounted for and leave no residual in the fitted model.

Lastly, after accounting for all point sources within the field of view with the *Fermi* tool gttsmap, one final refinement of the model file was performed, namely, the *Fermi* tool gtfindsrc which was used to determine a more precise localization of the source's RA and declination. The differences between the gtfindsrc results and the position found by DBSCAN all agree within the 95 per cent PSF and in most cases to better than 0°.1. The resulting positions, fluxes and TS values of all 28 DBSCAN clusters can be found in Table 2.

## 4 DISCUSSION

Using DBSCAN parameters EPS = 0°.4 and MinPts = 3 on 6.25 years of *Fermi*-LAT data for $E_\gamma \geq 100$ GeV, excluding data from $|b| < 10°$, we have found 49 sources which return a significant likelihood ratio. The positions of these sources can be seen in Fig. 2. Of the 61 extragalactic objects already existing in both the *Fermi*-LAT third point source catalogue (3FGL) and the TeVCat VHE catalogue (Table 3), 21 are also detected using DBSCAN. Of the remaining 28, 11 were found significant with follow up *Fermi* analysis (Table 2); 10 of these are in the 3FGL catalogue, which reports fluxes only up to 100 GeV.

### 4.1 DBSCAN performance

To estimate the performance of the DBSCAN algorithm in the case of VHE detections, we define the concept of *purity* as the number of sources with TS>25 (including the sources already in the TeVCat catalogue) against the total found by the DBSCAN clustering code. In Fig. 3(a) we show the number of sources with TS>25 and TS<25, along with the resultant purity, for all clusters found with DBSCAN using the range of investigated EPS values between the 68 per cent and 95 per cent PSF. As can be seen, the number of TS<25 sources found by DBSCAN rapidly increases for EPS>0°.3, while there is a marginal increase in the number of sources having TS > 25 above the same threshold. It should be noted however that the maximum number of significant sources found by DBSCAN occurs for EPS > 0°.4. As such, in order to maximize the number of sources with TS>25, with the maximum purity, an *EPS* of 0°.4 should be used by DBSCAN. For the remainder of the paper, we present our results based on the DBSCAN results with EPS = 0°.4.

To investigate the performance of the LRT significance, *s*, in equation 1, the LRT values for the clusters were compared to the TS values obtained with the *Fermi* Likelihood analysis. In Fig. 3(b), the LRT versus TS parameter space shows a clear correlation, with a large amount of quantization of the LRT distribution for low *s* values. This quantization is primarily due to the lack of background events detected with the LAT detector in $E_\gamma > 100$ GeV energy regime. While this suggests that the use of the LRT to define a DBSCAN cluster as significant results in a large number of false-positive detections, we note that our use of an LRT selection criteria of *s* > 2.0 is a conservative cut so as to guarantee the selection of all VHE sources in our sample. As such, while our use of *s* > 2.0 is sub-optimal for selecting VHE candidates with a high purity, Fig. 3(b) shows that this allows us to find all VHE sources present within our data set and thus maximises the number of new sources discovered. None the less, further work should be performed in order to investigate viable alternatives to the LRT that simultaneously maximises both the VHE-detection efficiency and the sample purity.

A full understanding of the efficiency of DBSCAN in this application is somewhat more complex, requiring detailed simulations and modelling of the *Fermi* VHE sky, which goes beyond the scope of this paper. However, estimations of DBSCAN efficiency can be found in Tramacere & Vecchio (2013) where, by simulating a range of false sky maps, they find it possible to achieve efficiencies of up to 96 per cent. This must be treated as an optimistic scenario as it is based on an optimal scan of the EPS–MinPts parameter space. We expect the efficiency to be much lower in our case due to our assumption of minimal background variation, which will be addressed in future work.

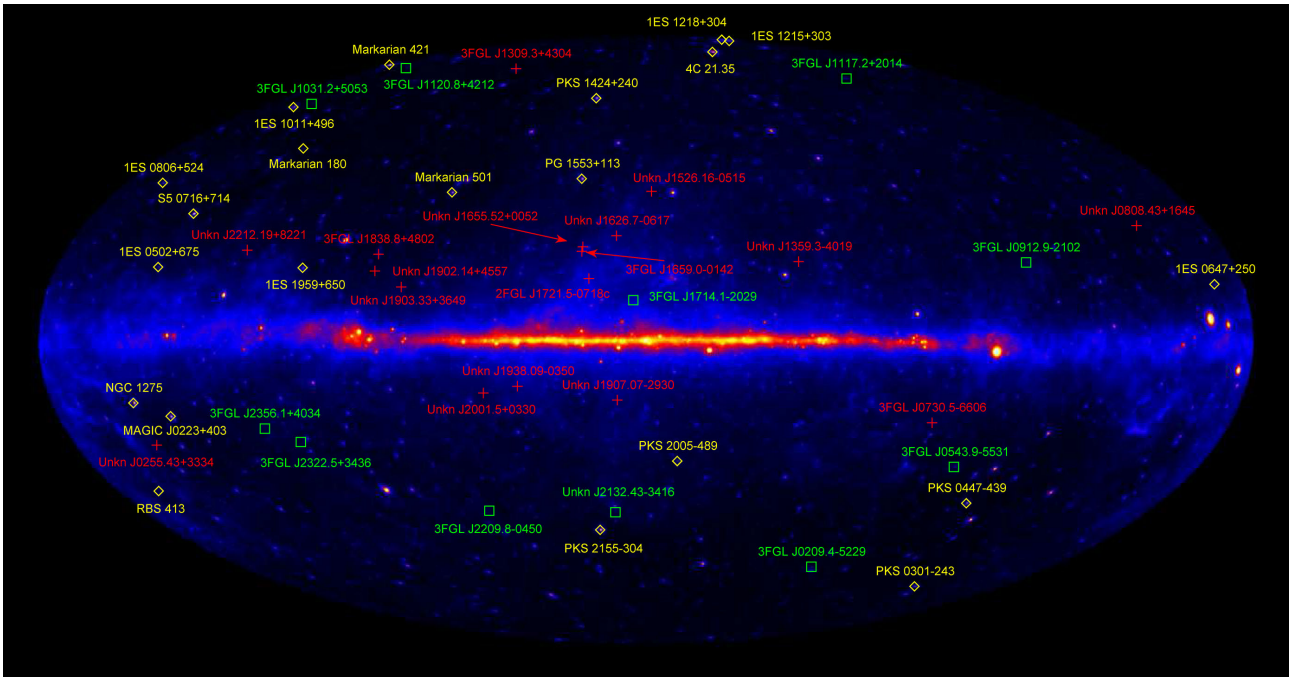

**Figure 2.** All sky image showing the locations of the clusters with TS > 25 (bold square), TS < 25 (cross) and the 21 objects that are already known TeVCat sources (diamond).

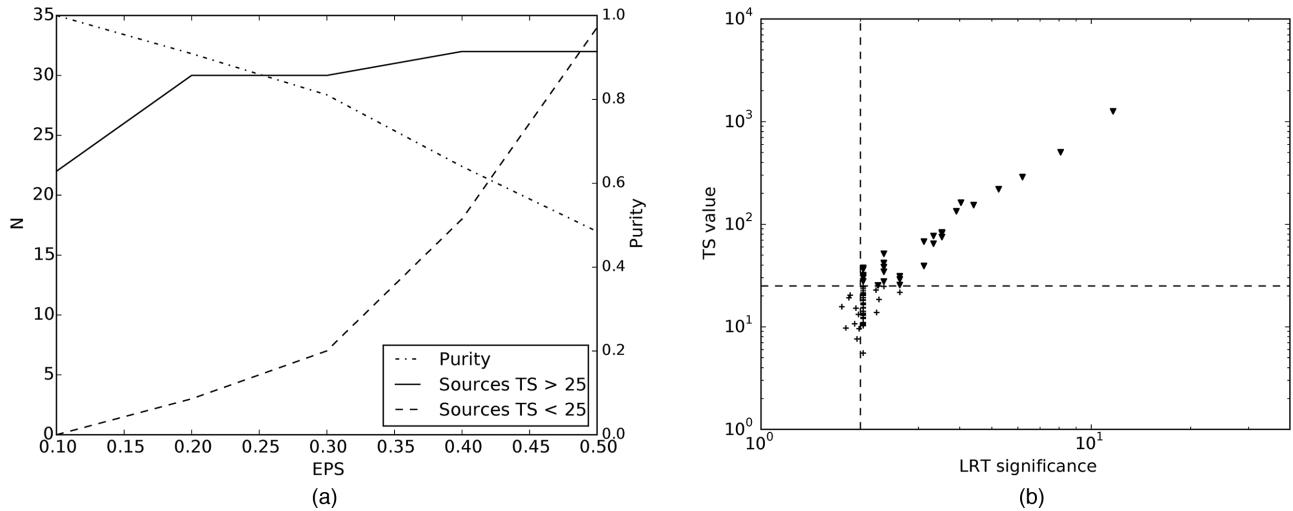

**Figure 3.** (a) Here we can see the effect of different values of EPS on the number of sources with TS <25 and TS >25. Once over 0°.2 the number of significant sources does not greatly increase until 0°.4 when one further source is added. However, the number of 'sources' that are not significant continually increases. (b) Comparing the value of LRT and TS for each cluster, we can see the quantization of the LRT due to a breakdown in the assumption that the number of signal and background events are not too small is clear. The solid triangles indicate the clusters with TS>25, while the crosses indicate the clusters with TS<25. The vertical dashed line indicates our LRT cut value, while the horizontal dashed line indicates our TS>25 cut value.

Although we note that there are still improvements to be made with the DBSCAN method, we draw attention to its capability of performing a quick, unbiased scan for potential 'seed' sources in the VHE *Fermi*-LAT sky which in this study has led to the detection of nine new VHE sources.

### 4.2 Detected VHE sources

To investigate the global properties of the *Fermi*-LAT VHE sources detected by the DBSCAN algorithm, we ran a binned likelihood analysis over the energy range 100 MeV to 100 GeV in order toobtain a reliable model file and fit for equation (2) with higher statistics. The data reduction method for this was the same as described in Section 3, but this time using an ROI of 12° centred on the published location of the source, keeping all modelled source parameters within this ROI free and freezing sources within an annulus 12° to 22° from the source of interest. For the analysis, the data were separated into 30 equally-spaced logarithmic energy bins. The resulting fluxes, spectral indices and TS values of the likelihood fits for these objects can be found in Table 2. For sources with TS <25, upper limits where calculated from the final fit and no spectral index is quoted.

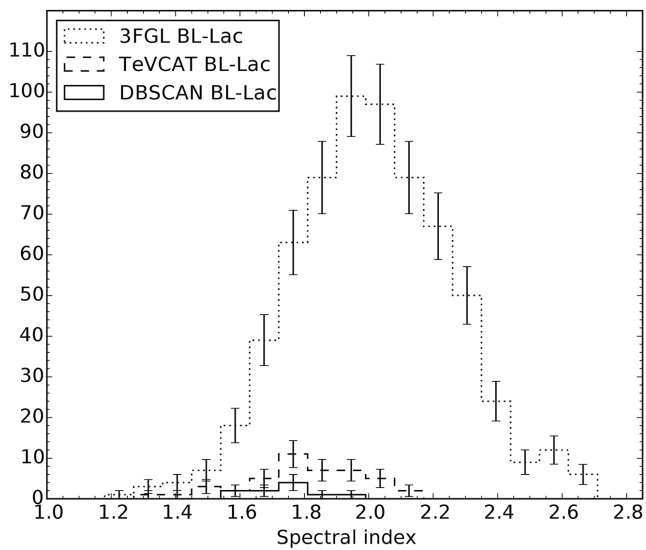

**Figure 4.** Histogram showing the spectral index distribution of the 3FGL and TeVCat BL Lac populations compared to those found in this work. Performing a standard independent 2-sample *t*-test infers that the 11 significant sources in Table 2 come from the same distribution as the VHE TeVCat sources.

Out of the 11 sources detected, we note that nine of them are blazars and all, except for 3FGL J1714.1-2029 which is of unknown AGN type, belong to the BL Lac class. The remaining two do not have any assigned counterparts. For each source we looked for temporal coincidence of the VHE events but found no evidence to suggest that the VHE photons originated in a single event.

The source of unknown type, 3FGL J2209.8-0450, which is a new addition since the 2FGL, is only 54.55 arcsec away from the radio source NVSS J220941-045111 (which is also connected to the X-ray object 1RXS J220942.1-045120). The second unassociated source has no known counterpart in the 3FGL (the closest known 3FGL source is the pulsar PSR J2124-3358, at $1°\!.69$ from the source) and no clear radio association, although its position is coincident with the galaxy group ESO 403-6. Although this source was detected in the 100 to 300 GeV range with a flux of $(2.45 \pm 1.84) \times 10^{-11}$ ph cm$^{-2}$ s$^{-1}$ it appears to have no significant emission in the energy range of 100 MeV to 100 GeV, making this an interesting VHE dark source. More work needs to be carried out in order to correctly identify counterparts for these sources.

In order to determine the likelihood that any of the unassociated sources with TS <25 are unresolved AGN, we checked for any coincidence with BZCAT sources (Massaro et al. 2009). We find no evidence of any association within the 95 per cent PSF, suggesting that a large proportion of these clusters arise from fluctuations in the background or from a larger unresolved structure.

As a first check of these results, we compared the spectral index found for each of our sources to those published in the 3FGL to look for any change over the last 4 years. We see no evidence of spectral hardening/softening, with the values agreeing within errors.

We then compared the spectral index distribution of the sources found using DBSCAN with the total 3FGL BL Lac population and those which also have ground-based VHE detections. The result of this comparison is shown in Fig. 4. In order to test whether the different distributions have the same mean and variance, we performed a standard independent 2-sample *t*-test on the DBSCAN sample and each of the spectral index distributions. Having initially set a significance level of 5 per cent, we find that the *Fermi* VHE sources detected with DBSCAN are better represented by the TeVCat BL Lacs, with a *P*-value of 0.368, than the total 3FGL BL Lacs for which we obtain a *P* value of 0.000 547.

We suggest that the sources we have detected with VHE emission, provided there are no spectral cut-offs, should be within reach of current and future ground based IACTs and should undergo follow-up observations.

## 5 CONCLUSIONS

We have presented an application of the clustering algorithm DBSCAN to 6.25 years of *Fermi*-LAT extragalactic data above 100 GeV, finding 49 clusters which were found significant using a LRT. Of the 28 which are not already known as VHE emitters in the TeVCat ground-based catalogue, we found 11 that were significant (TS>25) with follow up *Fermi* likelihood analysis. With the two sources RBS 0679 and RBS 0970 having previously been detected at $E \geq 100$ GeV (Brown 2014; Brown et al. 2014), we therefore present nine new VHE objects consisting of seven AGN and two unassociated sources.

We have performed a preliminary analysis into some of the global properties of these new *Fermi* VHE sources. Concerning the spectral indices derived from a fit between 100 MeV and 100 GeV, we see that these sources are more similar to the TeVCat BL Lac sources than to the overall 3FGL BL Lac population. We take this as a strong indication that these should be observable by current and future ground-based IACTs. A full analysis and description of these sources will be presented in future work.

## ACKNOWLEDGEMENTS

TPA would like to acknowledge the support of a studentship from the UK Science and Technology Facilities Council grant ST/K501979/1. AMB would like to acknowledge the financial support of Durham University. This work has made use of publicly available *Fermi*-LAT data from the High Energy Astrophysics Science Archive Research Center (HEASARC), provided by NASAs Goddard Space Flight Center. We would therefore like to acknowledge the *Fermi* collaboration for its tools and wealth of data. Finally, we would like to thank the reviewer for their comments and for helping improve the quality of this paper.

## SUPPORTING INFORMATION

Additional Supporting Information may be found in the online version of this article:

(http://mnras.oxfordjournals.org/lookup/suppl/doi:10.1093/mnras/stv1398/-/DC1).

Please note: Oxford University Press are not responsible for the content or functionality of any supporting materials supplied by the authors. Any queries (other than missing material) should be directed to the corresponding author for the article.

This paper has been typeset from a T$_{\rm E}$X/L$^{\rm A}$T$_{\rm E}$X file prepared by the author.